\documentclass[10pt]{article}

\makeatletter

\@ifundefined{date}{}{\date{}}
\usepackage[letterpaper]{geometry}
\usepackage{hicss}
\usepackage{times}
\usepackage[none]{hyphenat}
\usepackage{url}
\usepackage{latexsym}
\usepackage{indentfirst}
\usepackage{graphicx}
\usepackage{cite}
\usepackage{amsmath}
\usepackage{amssymb}
\usepackage{amsthm}
\usepackage{algorithm}
\usepackage[noend]{algpseudocode}
\usepackage{enumitem}
\usepackage{fancybox}
\usepackage{tcolorbox}
\usepackage{multirow}
\usepackage{pifont}
\usepackage{xcolor}

\definecolor{niceblue}{rgb}{0, 0.5, 1.0}
\graphicspath{{images/}}
\usepackage{float}
\floatstyle{ruled}

\newfloat{model}{thp}{lop}
\floatname{model}{Model}

\newfloat{problem}{thp}{lop}
\floatname{problem}{Problem}

\newtheorem{lemma}{Lemma}
\newtheorem{remark}{Remark}

\usepackage{hyperref}

\hypersetup{
    colorlinks=true,
    linkcolor=niceblue,
    filecolor=magenta,      
    urlcolor=niceblue,
    citecolor=niceblue}

\title{GPU-Accelerated DCOPF using Gradient-Based Optimization}

 \author{Seide Saba Rafiei \\ 
   University of Vermont \\
  {\underline{srafiei@uvm.edu}}\And
Samuel Chevalier \\
  University of Vermont \\
  {\underline{schevali@uvm.edu}} \\ }
  
\makeatother

\begin{document}
\maketitle 
\begin{abstract}

DC Optimal Power Flow (DCOPF) is a key operational tool for power system operators, and it is embedded as a subproblem in many challenging optimization problems (e.g., line switching). However, traditional CPU-based solve routines (e.g., simplex) have saturated in speed and are hard to parallelize. This paper focuses on solving DCOPF problems using gradient-based routines on Graphics Processing Units (GPUs), which have massive parallelization capability. To formulate these problems, we pose a Lagrange dual associated with DCOPF (linear and quadratic cost curves), and then we explicitly solve the inner (primal) minimization problem with a dual norm. The resulting dual problem can be efficiently iterated using projected gradient ascent.
After solving the dual problem on both CPUs and GPUs to find tight lower bounds, we benchmark against Gurobi and MOSEK, comparing convergence speed and tightness on the IEEE 2000, 4601, and 10000 bus systems. We provide reliable and tight lower bounds for these problems with, at best, 5.4x speedup over a conventional solver.
\end{abstract}

\subsubsection*{Keywords:}
\\

GPUs, DCOPF, linear programming, quadratic programming, conic relaxation, Lagrange dual, gradient-based optimization.

\section{Introduction}

Optimal power flow (OPF) is a crucial tool for enabling successful power system operations.
It plays a significant role in ensuring the economic operation of
power systems by determining the optimal active power output of generators
\cite{Ren_2022}~\cite{Stott1980OPF}. More specifically, DC OPF is
a simplified version of the OPF problem. It assumes a lossless and
unity voltage transmission network, and it can be efficiently solved\cite{Frank2012}~\cite{Alsac1990LP}. DCOPF is the cornerstone for many complex power system optimization problems (e.g., transmission line switching, unit commitment, etc.).
A plethora of research has focused on different methods for solving the DCOPF problem~\cite{Yang2023}~\cite{Quantum}.
Examples of these methods include barrier-based and Interior-Point Methods (IPMs)~\cite{PMjl}, commonly used by solvers like Gurobi, MOSEK, and Ipopt, and dual or primal simplex algorithms. While these methods are highly mature, they are serially constrained and do not parallelize effectively~\cite{SWIRYDOWICZ2022102870}. Thus, parallel solving thousands of DCOPF-like problems in a Unit Commitment or Transmission Switching Branch-and-Bound search tree is impractical. 
While metaheuristics (e.g., Genetic Algorithms, Particle Swarm Optimization) do parallelize, they can also be very slow due to their stochastic nature\cite{Metah}, and they also generally lack convergence guarantees, making them less trustworthy for industry uses.


Modern Graphic Processing Units (GPUs) enable massive computational parallelization, with many opportunities for power system
computational problems~\cite{GPU-acc}. 
GPUs are designed to handle thousands of
threads simultaneously, which makes them ideal for parallel computations. If posed correctly,
large optimization problems (e.g., DCOPF) can exploit this parallelization. Ferigo et al.~in \cite{Ferigo} introduced a GPU-enabled compact genetic algorithm,
which demonstrated considerably accelerated optimization for 
large-scale problems with millions of variables. The dynamic optimization
programming model in \cite{OConnell2017ADP} deployed on GPUs demonstrated
significant runtime improvements while efficiently managing memory
constraints. GPUs can enable more iterations and more comprehensive
searches within reasonable time frames \cite{Berrajaa2023}.
They are also designed to deliver high performance per watt. This
makes them energy-efficient for heavy computational tasks compared
to CPUs and is beneficial for large-scale and long-running optimization
tasks like multiperiod OPF~\cite{Zheng2023}. However, many modern solvers'
algorithms are currently unsuitable for implementation on GPUs. The Simplex algorithm, for
instance, is generally unsuited for GPUs due to its serial nature
and dependency on pivot operations, which are challenging to parallelize
effectively\cite{Horen1985-vv}. Also, some IPM operations are difficult to
map to the GPU's parallel architecture. Namely, parallelization
of sparse matrix factorizations (e.g., Cholesky, LU), which are critical
steps in IPMs, are slow and can dominate the computation time\cite{IPMGPU}.

There is ongoing research on designing numerical solution methods that can exploit GPU architectures. Ryu et al.~proposed a distributed optimization algorithm using ADMM. The primal and dual update steps only require matrix operations that can be implemented on GPUs, and they were able to show enhancements in computational time over CPU-based implementations\cite{Minseok}.
Deb et al.~introduced a gradient-based optimizer for economic load dispatch and showed competitive performance against other metaheuristic algorithms\cite{Deb}. Li et al.~ proposed an adaptive gradient-based method for parameter identification in power distribution networks in \cite{Li2022}, highlighting a robust performance and low error rates compared to heuristic algorithms.~\cite{Shrirang:2023} used gradient-based routines from the Neuromancer library to solve a constrained optimization problem via constraint penalization. Similarly, ~\cite{chevalier2023parallelized} designed a parallelized, Adam-based solver for the AC unit commitment problem, where constraint penalization was aggressively leveraged.

All of these approaches utilize either alternating primal-dual update steps or direct constraint penalization in the objective function. Inspired by recent, gradient-based approaches for verifying the performance of large-scale neural networks~\cite{betacrown}, our paper adopts a new approach entirely: we formulate a Lagrange Dual associated with a DCOPF problem (with linear or quadratic cost terms), and then we explicitly solve the inner (primal) minimization using a dual norm. We push the primal constraints into the dual domain by applying the dual norm and linear/conic projections. The resulting dual problem can be solved via projected gradient ascent, which is perfectly suitable for GPU implementation and is naturally parallelizable and fast~\cite{GradPar}.


We then solve the dualized DCOPF problem using three variants of Gradient-Based Optimization (GBO), both on CPUs and GPUs, and we compare the results based on standard convergence metrics, bench-marking against Gurobi and MOSEK. The main challenges motivating this work and the corresponding contributions of this paper are explained below.
\begin{enumerate}
   
    \item Many uses of DCOPF (e.g., within a Branch-and-Bound search tree) need a valid solution cost bound. In this paper, we solve a dualized problem with constraints that can be satisfied via trivial projections at every gradient step.
    The main advantage of this method is that, by solving the dual with an iterative GBO approach, the process can be paused at any iteration and still provide a valid lower bound on the generation cost. 

    \item Naive primal and dual formulations of the DCOPF problem cannot be efficiently solved on GPUs. By formulating a dual with trivial constraint satisfaction (i.e., by projecting dual variables into the nonnegative orthant), we are able to load and solve large-scale DCOPF problems on a single GPU.


    \item Finally, many competitive gradient-based tools have emerged in the literature. We tested and analyzed the sensitivity of three different GBO 
    variants (Adam, AdaGrad, and Gradient with Momentum) for solving our formation, and we identify key hyperparameters and methods that yield swift and reliable convergence on three standard test cases.

    
\end{enumerate}
Notably, this paper focuses on solving singular DCOPF problems. However, our GPU-based approach allows us to parallel solve hundreds of such problems at a similar speed, which will be the foundation for future work. The rest of the paper is structured as follows. In Section \ref{sec: dcopf}, we review the structure of DCOPF. In Section \ref{sec: dual}, we explain the reformulations needed to pose the problem as Linear Programming (LP) and Quadratic Programming (QP) to adapt the dual problems to GBO.  Test results are given in Section \ref{sec: results}, and conclusions are drawn in Section \ref{sec:conclusion}.

\section{DCOPF Problem Formulation}\label{sec: dcopf}

\label{sec: Complete_Verification} 

For each bus $i$ and $j$ in a power system, assuming the voltages to be one (i.e. $V_{i}\approx1$
and $V_{j}\approx1$) and the phase differences across the transmission
lines $(\delta_{i}-\delta_{j})$ to be very small, we can represent the problem of DCOPF as minimizing the sum of the total cost of power generated by each generator with its associated cost $c_{g}$, subject
to constraints arising from power balance, line flow limits, and generation
limits as eq.~\eqref{eq:DCOPF} \cite{hesamzadeh2014chapter6}
\begin{subequations}\label{eq:DCOPF}
\begin{align}
\underset{p_{g}}{\min}\quad & c_{g}^{T}p_{g}\\
{\rm s.t.}\quad & 1^{T}p_{g}-1^{T}p_{d}=0\\
& \underline{p}_{f}\leq H_a({p}_g-{p}_d)\leq\overline{p}_{f}\label{eq: Ha}\\
& \underline{p}_{g}\leq p_{g}\leq\overline{p}_{g},
\end{align}
\end{subequations}
where  $H = \Omega {E} ( {E}^{T} \Omega {E})^{-1}$ is the power transfer distribution factor (PTDF) matrix, in which ${E} \in \mathbb R^{n_{l}\times n_{b}-1}$ is the reduced incidence matrix, and $\Omega={\rm diag}(b)\in{\mathbb R}^{n_l\times n_l}$ is the diagonal susceptance matrix. In \eqref{eq: Ha}, $H_a$ is the PTDF matrix augmented with a zero column associated with the slack bus.
Without loss of generality, the vector ${p}_g-{p}_d$ is the full net power injection vector at each bus. Notably,  this DCOPF problem can also accommodate quadratic generator cost curves by appending $p_g^T M p_g$ to the objective. In this case, the LP becomes a convex QP. The following sections describe how these L/QPs can be reformulated and solved in the dual space via gradient-based optimization.

\subsection{Canonicalizing the Problem}
In order to reformulate problem~\eqref{eq:DCOPF} into the dual space, we first canonicalize it. Recasting the power generation vector $p_{g}$ at each bus as the normalized decision variable $x$ (see appendix A.2), the problem can be transformed into~\eqref{eq:general}, which has the standard form of an LP (if $M=0$) or a QP otherwise:\footnote{DCOPF is the guiding application in this paper, but the methods shown can be used to solve any optimization problem of the same form numerically.}


\begin{subequations}
\begin{align}
\underset{||x||_{\infty}\leq1}\min\quad & x^{T}Mx+c^{T}x\\
{\rm s.t.}\quad & Ax-b=0\;\,:\; \lambda\\
 & Cx-d\le0\;:\; \mu.
\end{align}
\label{eq:general}
\end{subequations}
\section{The Dual Problem}\label{sec: dual}
To construct a dual problem that can be efficiently solved on GPUs, we first pose the dual problem, and then we solve the inner dualized primal problem with a dual norm. The Lagrangian of the primal problem \eqref{eq:general} is formed as
\begin{align}
\mathcal{L}=
 x^{T}Mx+c^{T}x+\lambda^{T}(Ax\!-\!b)+\mu^{T}(Cx\!-\!d),\label{eq:lagrangian}
\end{align}
 As shown in eq.~\eqref{eq:dual function}, the dual function finds the greatest lower bound on the Lagrangian for any $x$, given fixed values of $\lambda$ and $\mu$:
\begin{equation}
\mathcal{G}(\lambda,\mu)=\inf_{x}\;\mathcal{L}(x,\lambda,\mu).
\label{eq:dual function}
\end{equation}
Maximizing this function over the feasible set of dual variables yields the dual problem~\cite{boyd2004convex}:
\begin{equation}
\gamma=\max_{\mu\geq0,\lambda}\;\mathcal{G}(\lambda,\mu).
\label{eq:dual}
\end{equation}
Since the DCOPF problem is convex and generally obeys Slater's condition, the optimal value of the dual problem ($\gamma$) is equal to the DCOPF solution objective value. Furthermore, any set of feasible (not necessarily optimal) dual variables can be used to provide a valid lower bound to the optimal objective value~\cite{boyd2004convex}. This observation helps motivate solving the dual problem in this paper instead of directly solving the primal. Using an iterative method, the dual problem will give us a valid lower bound at any time we decide to stop iterating (assuming the dual variables are projected feasible).

\subsection{The Dual Problem of the LP}
Assuming $M=0$ in \eqref{eq:general}, the QP simplifies to an LP. Since the \textit{normalized} input generation variables are constrained to live within the infinity norm ball, we can solve the inner minimization directly with a dual norm (see appendix A.2), as in \cite{betacrown}:
\begin{subequations}
\begin{align}\gamma_{l} & =\max_{\mu\geq0,\lambda}\min_{\|x\|_{\infty}\leq1}\;{\mathcal{L}}(x,\lambda,\mu)\\
 & =\max_{\mu\geq0,\lambda}\underbrace{-\|c^{1}+\lambda^{T}A+\mu^{T}C\|_{1}\!-\!\lambda^{T}b\!-\!\mu^{T}d}_{g_l(\mu,\lambda)},\label{eq:LP dual}
\end{align}
\end{subequations}
where \eqref{eq:LP dual} is a maximization problem with dual feasibility as the only constraint. 
\begin{remark}
   For any set of feasible $\lambda$ and $\mu \ge 0$ variable values, $g_l(\mu,\lambda)$ provides a valid lower bound on the optimal objective $\gamma_l$:
   \begin{align}
\gamma_{l}\ge g_l(\mu,\lambda),\quad\forall\mu\ge0,\lambda.
\end{align}

\end{remark}

\subsection{The Dual Problem of the QP}
We now consider the more involved QP of \eqref{eq:general}, i.e., where the diagonal matrix $M$ has some nonzero elements.
We introduce an auxiliary variable $t$ to bound the quadratic term and rewrite the problem as \eqref{eq:socp}, which is a convex Second Order Cone Program (SOCP):
\begin{subequations}\label{eq:socp}
\begin{align}
\min_{x}\quad & t+c^{T}x\\
{\rm s.t.}\quad & 
x^{T}Mx\le t\;\;\,:\; s\\
 & Ax-b\le0\;\,:\; \lambda\\
 & Cx-d=0\;\,:\; \mu.
\end{align}
\end{subequations}
To obtain the dual of this program, we note that the dual variable associated with the SOC constraint must always belong to the dual cone (see appendix A.3)\cite{MOSEK}. Dualizing the equality and inequality constraints, the SOCP problem is reformulated as
\begin{subequations}
\label{eq:socp-1} 
\begin{align}
\gamma_q = \max_{\mu \geq 0, \lambda} \; \min_{x} \quad & t + c^{T}x + \lambda^{T}(Ax - b) \nonumber\\
& \quad\quad\quad\; + \mu^{T}(Cx - d) \label{eq:sub1} \\
\text{s.t.} \quad & (t, 0.5, \sqrt{M}x) \in K^{\text{RSOC}},
\end{align}
\end{subequations}
where $\sqrt{M}$ is the element-wise square root of all matrix entries, and $K^{\text{RSOC}}$ is a rotated second-order cone constraint corresponding to 
\begin{align}
2\cdot 0.5 \cdot t\ge M_{1,1}x_{1}^{2}+\cdots+M_{n,n}x_{n}^{2}.
\end{align}
Dualizing this SOC constraint~\cite{MOSEK} leads to
\begin{align}
\gamma_{q}=\max_{{\tilde s}\in K_{*}^{{\rm RSOC}},\mu\ge0,\lambda}&\;\min_{x}\quad  t+c^{T}x+\lambda^{T}\left(Ax-b\right)+ \nonumber \\
\mu^{T}\!\left(Cx-d\right) &\! -\! (s_{1}t+0.5s_{2}\!+x^{T}\sqrt{M}s),
\label{eq:socp_dual}
\end{align}
where ${\tilde s}=(s_{1},s_{2},s)$, and where  ${\tilde s}\in K_{*}^{{\rm RSOC}}$ corresponds to dual rotated second-order cone constraint
\begin{align}
    2s_{1}s_{2}\ge s^{T}s.
\end{align}
We now augment the problem and solve it with the dual norm. To do so, we define $\hat{s}=({s_{1},s})$ and $\hat{x}=(t,x)$, with corresponding matrices\footnote{For notational simplicity, we do not explicitly define these matrices, nor do we state the normalization of primal variable $t$, but these steps allow us to canonicalize the problem.} $\hat{c}$, $\hat{M}$, $\hat{A}$, and $\hat{C}$, such that
\begin{align}\label{eq: pre-dual_n}
\gamma_{q}=\max_{{\tilde s}\in K_{*}^{{\rm RSOC}},\mu\ge0,\lambda}\;\min_{\left\Vert \hat{x}\right\Vert _{\infty}\le1}&\quad  \hat{c}^{T}\hat{x}+\lambda^{T}\left(\hat{A}\hat{x}-b\right)+
\nonumber \\
 \mu^{T}\left(\hat{C}\hat{x}-d\right)-&\hat{s}^{T}\sqrt{\hat{M}}\hat{x}-0.5s_{2}.
\end{align}
To apply the dual norm to the inner minimization of \eqref{eq: pre-dual_n}, we need to explicitly bound the additional primal variable $t$, which is the slack variable which upper bounds the SOC constraint (i.e., upper bounds the sum of the generators' quadratic costs). We denote this bound as $\overline t$, which we safely estimate as the maximum sum across all generators' quadratic costs.

Using the dual norm to minimize over $B_{\infty}$, we derive the dual problem as
\begin{align}\label{eq:dn_QP}
\gamma_{q}=\!\!\!\!\max_{{\tilde s}\in K_{*}^{{\rm RSOC}},\mu\ge0,\lambda}-&\left\Vert \hat{c}^{T}+\lambda^{T}\hat{A}+\mu^{T}\hat{C}-\hat{s}^{T}\sqrt{\hat{M}}\right\Vert _{1} 
\nonumber \\
 &-0.5s_{2}-\lambda^{T}b-\mu^{T}d.
\end{align}
While it seems we would need to continuously project ${\tilde s}$ into the dual cone $K_{*}^{{\rm RSOC}}$ to solve \eqref{eq:dn_QP} with a gradient-based method, this is interestingly not the case. To explain why, we note that $s_2$ only shows up as a linear term in the \eqref{eq:dn_QP}, i.e., it does not appear in the 1-norm. Therefore, at optimality, the dual SOC constraint will necessarily hold with equality ($s_{1}s_{2}= s^{T}s$) to ensure that the primal SOC constraint is satisfied. Therefore,
\begin{align}\label{eq: s2}
s_{2} & =\frac{s^{T}s}{2s_{1}}.
\end{align}
Thus, we replace $s_2$ in eq.~\eqref{eq:dn_QP} with \eqref{eq: s2}. Finally, we note that if we isolate terms in the objective involving $s_1$, we get
$-|\overline{t}-s_{1}\overline{t}|-0.25{s^{T}s}/s_1$, where $\overline{t}$ is the upper bound on $t$. Since $\overline{t}\ge 0.25s^{T}s$, using a super-gradient argument, we have $s_1=1$ as the maximizer. Therefore, we obtain the final dual problem for the case of quadratic cost curves:
\begin{align}\label{eq:QP_finalgamma}
\gamma_{q}&=\max_{\mu\ge0,\lambda,s}\quad-\left\Vert {c}^{T}+\lambda^{T}{A}+\mu^{T}{C}-{s}^{T}\sqrt{{M}}\right\Vert _{1}
\nonumber \\
&\quad\quad\quad\quad\quad\,\, -0.25s^{T}s-\lambda^{T}b-\mu^{T}d\\
&= \max_{\mu\ge0,\lambda,s}\quad g_q(\mu, \lambda, s),
\end{align}
where the vector $s$ is a set of unconstrained dual variables, and all $\hat{\cdot}$ terms have been removed, since $s_1=1$ is implied. Numerical testing shows that the objective solutions of \eqref{eq:socp} and \eqref{eq:QP_finalgamma} exactly agree.
\begin{remark}
   For any set of feasible $\mu \ge 0$, $\lambda$, and $s$ variable values, $g_q(\mu,\lambda, s)$ provides a valid lower bound on the optimal objective $\gamma_q$:
   \begin{align}
\gamma_{q}\ge g_q(\mu,\lambda),\quad\forall\mu\ge0,\lambda, s.
\end{align}
\end{remark}

\subsection{Using GBO Variants to Solve the Dual}
In order to solve \eqref{eq:LP dual} (in the LP case) and \eqref{eq:QP_finalgamma} (in the QP case), we use projected gradient ascent, computed primarily on GPUs. While there are many gradient-based routines in the literature, in this paper, we focus on three successful variants of gradient-based optimization: Adaptive Moment Estimation (Adam)\cite {Adam}, Adaptive Gradient Algorithm (AdaGrad) \cite{Adagrad}, and Gradient with Momentum (GDM) \cite{GDM}. These methods are similar in structure but different in the step size update rules. As an example, algorithm 1 shows the steps to GDM.

\begin{algorithm}
\caption{Gradient with Momentum}
\label{algo:gradient_momentum}
{\small\textbf{Require:} Initial $\lambda = 0$, $\mu = 0$, $v_{\lambda} = 0$, $v_{\mu} = \mathbf{0}$, step size $\alpha$, momentum.\\
\textbf{Ensure:} Optimal $\lambda$ and $\mu$.

\begin{algorithmic}[1]
\For{each iteration $i$ from 1 to $n_{its}$}
    \State $\nabla_{\lambda}\gamma \gets -\sum_{j}\text{sign}({c_{j}} + \lambda_{j}{A}_{j} + \mu_{j}{C}_{r_{j}}){A}_{j} + {b}$
    \State $\nabla_{\mu}\gamma \gets -\sum_{j}\text{sign}({c_{j}} + \lambda_{j}{A}_{j} + \mu_{j}{C}_{r_{j}}){C}_{r_{j}} + {d}_{r}$
    \State $v_{\lambda} \gets \text{Momentum} \times v_{\lambda} + \alpha \nabla_{\lambda}\gamma$
    \State $v_{\mu} \gets \text{Momentum} \times v_{\mu} + \alpha \nabla_{\mu}\gamma$
    \State Velocity Update:\\
    \qquad \qquad$\lambda \gets \lambda + v_{\lambda}$,
  \quad $\mu \gets \mu + v_{\mu}$
    \State Gradient Update:\\
   \qquad \qquad $\lambda \gets \lambda + \alpha \nabla_{\lambda}\gamma$,\quad $\mu \gets \mu + \alpha \nabla_{\mu}\gamma$
    \State $\mu \gets \max(\mu, 0)$
    \State $\gamma^{(i+1)} \gets
    -\|c + \lambda^{T}{A} + \mu^{T}{C}_{r}\|_{1} - \lambda^{T}{b} -\mu^{T}{d}_{r}$
    \If{$|\gamma^{(i+1)} - \gamma^{(i)}| < \text{tolerance}$}
        \State \textbf{Stop}
    \EndIf
\EndFor
\end{algorithmic}}
\end{algorithm}

\section{Test Results}\label{sec: results}

In this section, we present the results attained from solving the problem of DCOPF via GBO on CPU and GPU. We compare the performance of the three GBO variants on the LP with regard to iterations and speed. The tests are done on the IEEE 2000, 4601, and 10000 bus systems separately~\cite{Babaeinejadsarookolaee:2019}. Then, we investigate the effectiveness of using GPUs for solving the two (relatively small and big) problems. At the end, we present an analysis of the hyperparameters associated with the GBO routines and how the tuning process can affect the convergence. All the simulations are run on Apple M2 Pro with 32 GB of shared memory between CPU and GPU, with a total of 12 cores of CPU and 19 cores of GPU and Metal 3 support via Metal.jl~\cite{Besard_Metal_jl_2022}. All code is written with minimal memory allocation. Benchmarks were solved with Gurobi and MOSEK 10.

\subsection{Solving the LP with GBO}
In the case of the 2k-bus system, AdaGrad and GDM  provide a tight solution bound in almost half the best benchmark time (Gurobi), with gaps of 1.09\% and almost zero, respectively. These results cannot necessarily be generalized to any system. Therefore, two other tests are done on the 4601-bus and the 10k-bus system. In the 4601 system, GDM provides a solution of 4.02e-3 gap in almost one-forth of the best benchmark whereas for the 10k system, Adam provides a lower bound with a gap of 0.2\% in almost one-fifth the time of the best benchmark. Thus, as the size of the system grows, Adam seems to be more effective, while for the smaller system, GDM provides a faster and smoother convergence.

\begin{table}[ht]
\centering 
\caption{Comparison of CPU and GPU LP solver performance for different algorithms (2k-bus).}
\label{table:2000_LP}
\begin{tabular}{|c|c|c|c|c|}
\hline
Algorithm & t(s)-CPU & t(s)-GPU & It & Gap(\%) \\
\hline
Adam & 0.39 & 1.40 & 125 & 0.07 \\
\hline
AdaGrad & 0.08 & 0.70 & 121 & 1.09 \\
\hline
GDM & 0.10 & 0.74 & 62 & 4e-4 \\
\hline
Gurobi & 0.21 & - & 23 & - \\
\hline
MOSEK & 0.65 & - & 10 & - \\
\hline
\end{tabular}
\end{table}

\begin{table}[ht]
\centering 
\caption{Comparison of CPU and GPU LP solver performance for different algorithms (4601-bus).}
\label{table:4601_LP}
\begin{tabular}{|c|c|c|c|c|}
\hline
Algorithm & t(s)-CPU & t(s)-GPU & It & Gap(\%) \\
\hline
Adam & 0.37 & 0.61 & 45 & 0.1 \\
\hline
AdaGrad & 0.23 & 0.53 & 56 & 6.17e-4 \\
\hline
GDM & 0.15 & 0.36 & 28 & 4.02e-3 \\
\hline
Gurobi & 0.63 & - & 78 & - \\
\hline
MOSEK & 1.37 & - & 17 & - \\
\hline
\end{tabular}
\end{table}

\begin{table}[ht]
\centering 
\caption{Comparison of CPU and GPU LP solver performance for different algorithms (10k-bus).}
\label{table:10000_LP}
\begin{tabular}{|c|c|c|c|c|}
\hline
Algorithm & t(s)-CPU & t(s)-GPU & It & Gap(\%) \\
\hline
Adam & 2.25 & 1.85 & 77 & 0.2 \\
\hline
AdaGrad & 8.60 & 6.40 & 269 & 0.30 \\
\hline
GDM & 15.20 & 13.90 & 28 & 0.16 \\
\hline
Gurobi & 10.06 & - & 823 & - \\
\hline
MOSEK & 11.07 & - & 21 & - \\
\hline
\end{tabular}
\end{table}

Comparing tables \ref{table:2000_LP}, \ref{table:4601_LP}, and \ref{table:10000_LP}, it is worth noting that for the smaller systems, since the solve time is small, the overhead caused by data transfer between CPU and GPU is dominant, so the GPU solve time is slower than the parallelized CPU solve time. However, this is not the case for the large system, and we can observe that the iterations run faster on the GPU as the size of the system grows. 

 Using a step decay to decrease the step size in each iteration has proven to be effective in obtaining a faster GBO convergence in several sources\cite{lr}. In this case, interestingly enough, using the step decay for GDM on a 10000 bus system will either result in a higher solution gap or more solve time. The decision to use step size decay for LP depends on whether higher precision or speed is prioritized. 
 
\subsection{Solving the QP with GBO}
Looking at the results, GDM gives the fastest and tightest lower bound on the QP problem in the 2k-bus system, while Adam is doing better in the QP problem of the 10k system. The strength of this method is perfectly observable from tables \ref{table:QP_10000} and \ref{table:10000_LP} as we present lower bounds with tightness of 0.2\% and 0.44\% for the 10k system in half the solve time of the best benchmark.

\begin{figure}[!tb]
\centering \includegraphics[width=1\linewidth]{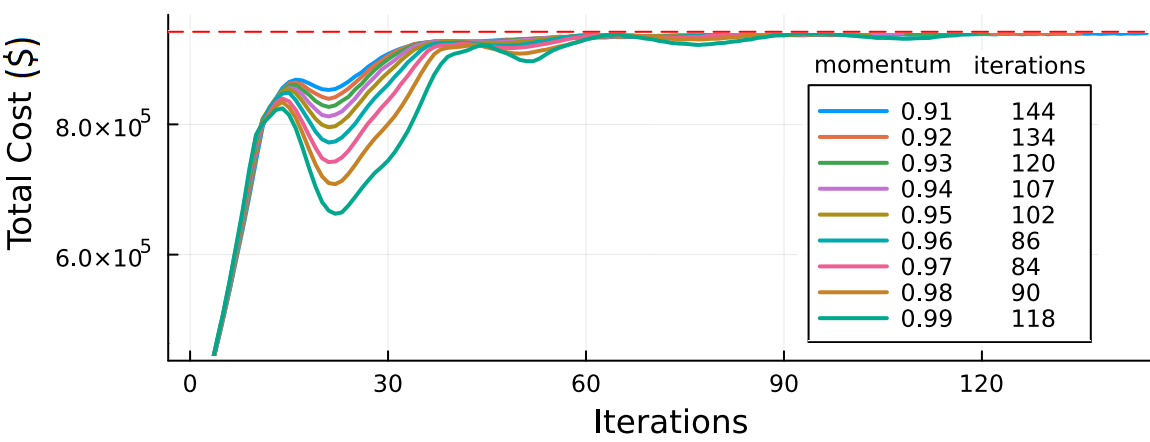}
\caption{GDM sensitivity with the change of momentum, $\alpha$=0.15.}
\label{fig: GDM-a0.15} 
\end{figure}

\begin{figure*}
\centering \includegraphics[width=1\linewidth]{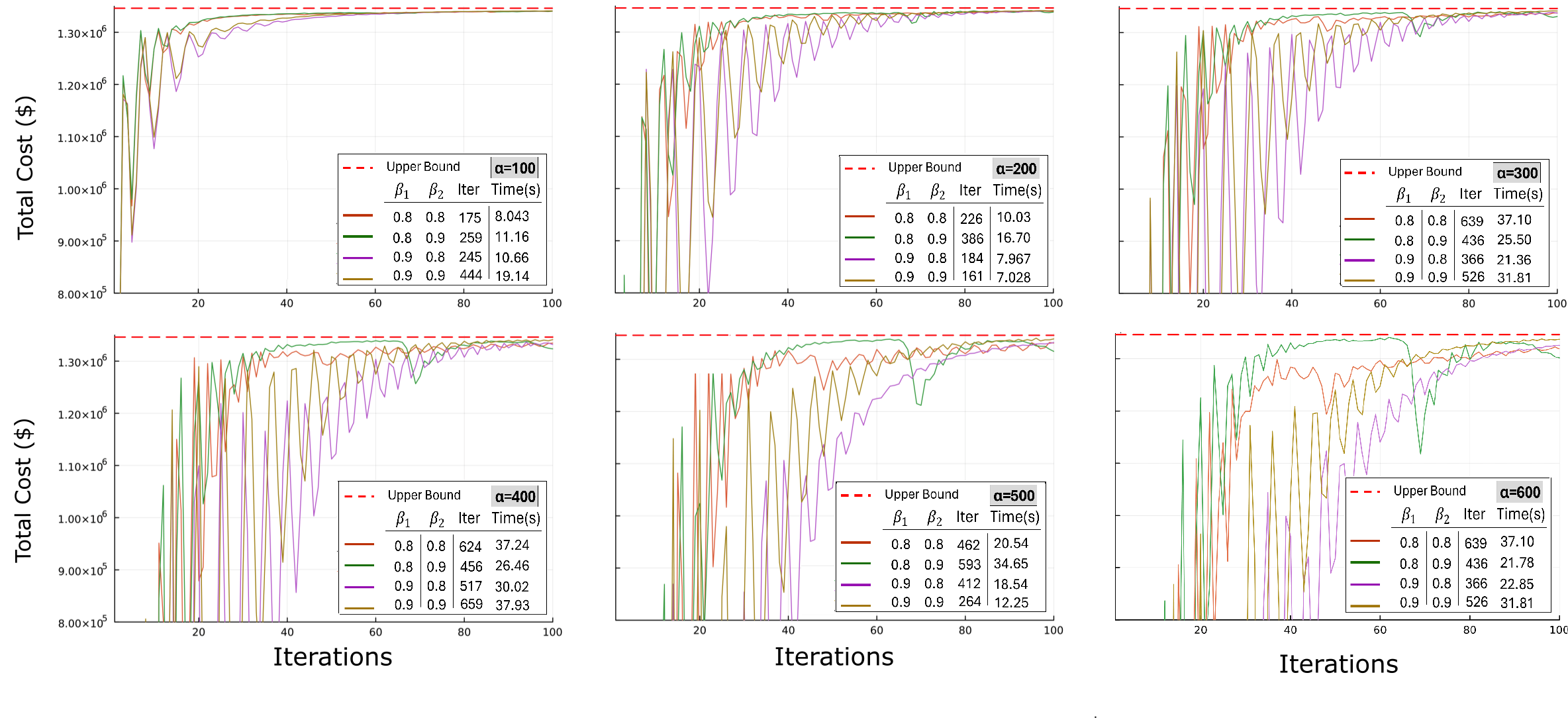}
\caption{Dual objective sensitivity analysis of Adam with change of $\alpha$, $\beta_{1}$, and $\beta_{2}$, QP, 10k-bus system.}
\label{fig: adam} 
\end{figure*}

We benchmark the QP against MOSEK and Gurobi. Tables \ref{table:QP_2000}, \ref {table:QP_4601}, and \ref{table:QP_10000} show the results for solving the QP on a 2000, 4601, and 10000 bus systems, respectively. The advantages of using the SOCP dual cone projection are notable. As opposed to the GDM method in LP, using step decay showed promising results for the 10k and 4601-bus case for the QP. It enabled faster and more stable convergence in the end phase.
The process of tuning the hyper-parameters is very important and greatly affects the convergence of GBO. As can be seen from Fig.~\ref{fig: GDM-a0.15}, for the case of GDM, as we increase momentum, the rise time decreases, but the stability of the convergence is compromised. A balance should be set, considering the priority of the problem solution. Note also how the stability in the first phase and the iterations change for Adam as $\alpha$, $\beta_{1}$ and $\beta_{2}$ change in Fig.~\ref{fig: adam}. The tuned hyperparameters for all the results in this paper are presented in tables \ref{table:hyperparameters_2k},~\ref{table:hyperparameters_4601}, and \ref{table:hyperparameters_10k}.
\begin{table}[!bt]
\centering 
\caption{LP and QP Hyperparameter settings for different algorithms (2k-bus).}
\label{table:hyperparameters_2k}
\begin{tabular}{|c|c|c|c|c|c|}
\hline
 & Algorithm & $\alpha$ & $\rho$ & $\beta_1$ & $\beta_2$ \\
\hline
\multirow{3}{*}{LP} & Adam & 500 & - & 0.9 & 0.997 \\
\cline{2-6}
 & Adagrad & 1500 & - & - & - \\
\cline{2-6}
 & GDM & 0.601 & 0.947 & - & - \\
\hline
\multirow{3}{*}{QP} & Adam & 500 & - & 0.9 & 0.99 \\
\cline{2-6}
 & AdaGrad & 500 & - & - & - \\
\cline{2-6}
 & GDM & 0.9 & 0.9 & - & - \\
\hline
\end{tabular}
\end{table}

\begin{table}[!bt]
\centering 
\caption{LP and QP Hyperparameter settings for different algorithms (4601-bus).}
\label{table:hyperparameters_4601}
\begin{tabular}{|c|c|c|c|c|c|}
\hline
 & Algorithm & $\alpha$ & $\rho$ & $\beta_1$ & $\beta_2$ \\
\hline
\multirow{3}{*}{LP} & Adam & 485 & - & 0.90 & 0.96 \\
\cline{2-6}
 & Adagrad & 2725 & - & - & - \\
\cline{2-6}
 & GDM & 6.5 & 0.94 & - & - \\
\hline
\multirow{3}{*}{QP} & Adam & 750 & - & 0.5 & 0.999999 \\
\cline{2-6}
 & AdaGrad & 2250 & - & - & - \\
\cline{2-6}
 & GDM & 2.21 & 0.8 & - & - \\
\hline
\end{tabular}
\end{table}

\begin{table}[!bt]
{
\centering 
\caption{LP and QP Hyperparameter settings for different algorithms (10k-bus).}
\label{table:hyperparameters_10k}
\begin{tabular}{|c|c|c|c|c|c|}
\hline
 & Algorithm & $\alpha$ & $\rho$ & $\beta_{1}$ & $\beta_{2}$ \\
\hline
\multirow{3}{*}{LP} & Adam & 496 & - & 0.9 & 0.991 \\
\cline{2-6}
 & AdaGrad & 420 & - & - & - \\
\cline{2-6}
 & GDM & 0.9 & 0.99 & - & - \\
\hline
\multirow{3}{*}{QP} & Adam & 500 & - & 0.9 & 0.997 \\
\cline{2-6}
 & AdaGrad & 1500 & - & - & - \\
\cline{2-6}
 & GDM & 0.99 & 0.99 & - & - \\
\hline
\end{tabular}}
\end{table}

\begin{figure}[!tb]
\centering \includegraphics[width=1\linewidth]{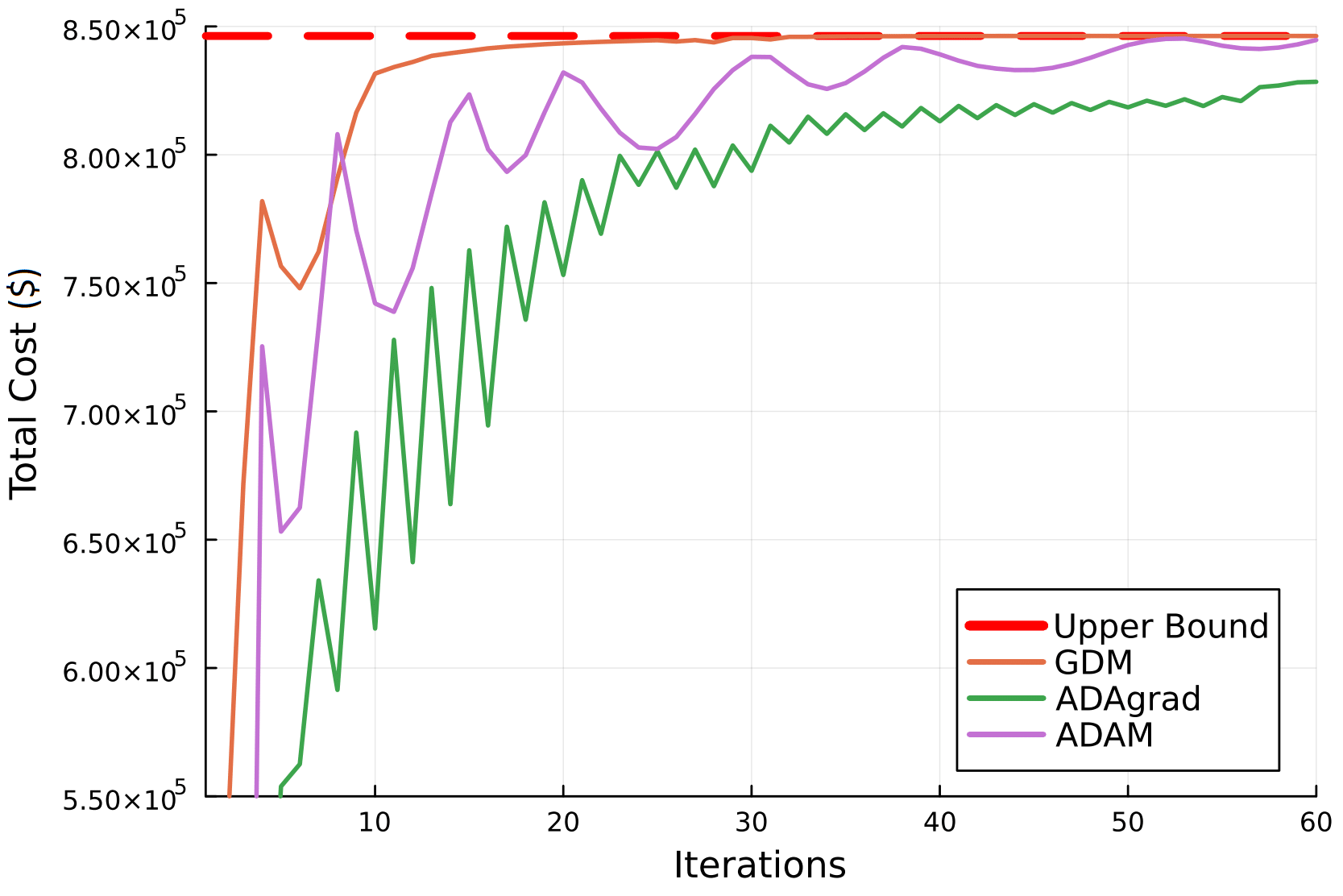}
\caption{Dual solution behavior of GBO, LP, 2k-bus system.}
\label{fig: LP-2000} 
\end{figure}
\begin{figure}[!tb]
\centering \includegraphics[width=1\linewidth]{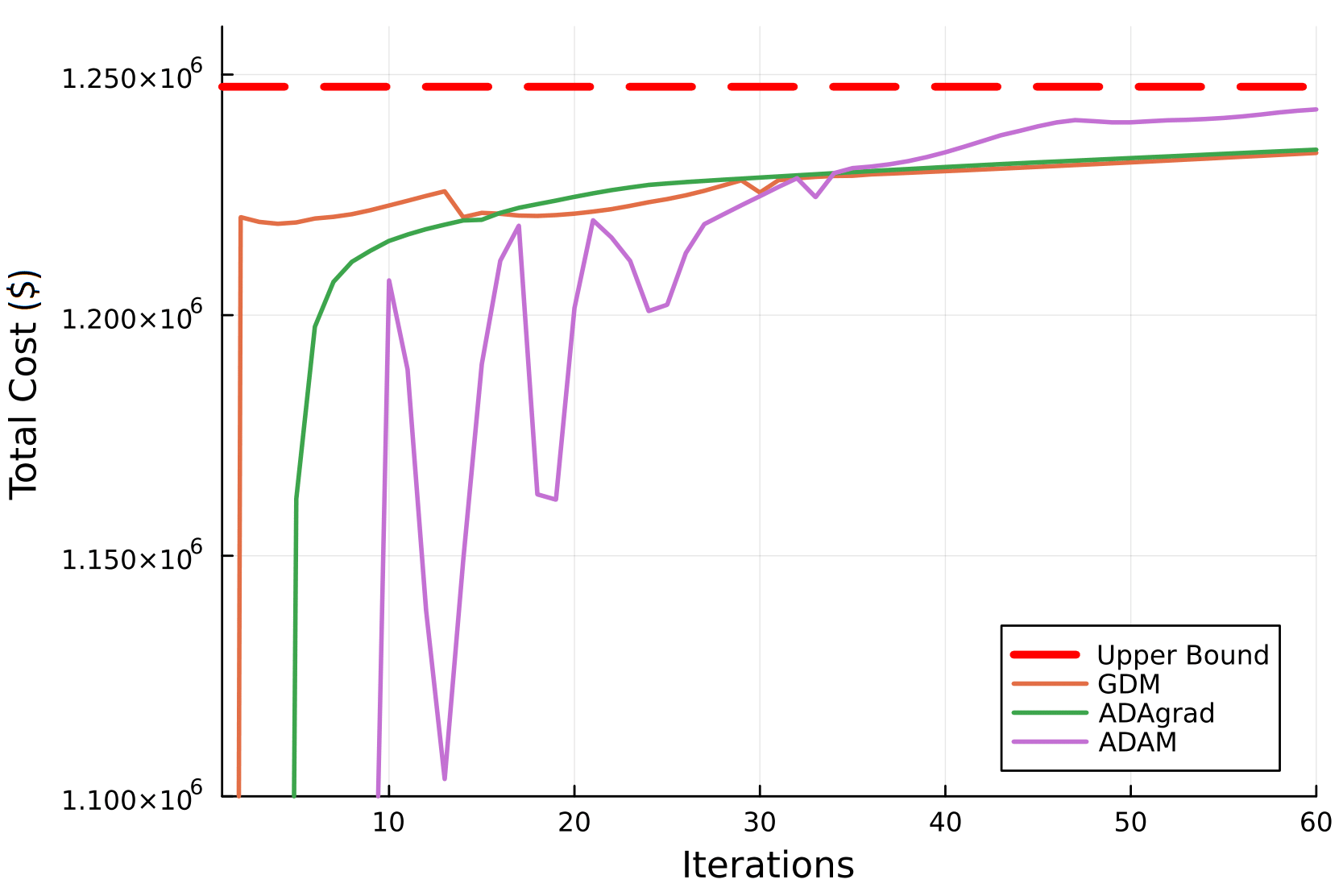}
\caption{Dual objective behavior of GBO, LP, (10k-bus) system.}
\label{fig: LP-10000} 
\end{figure}

\begin{table}[!bt]
\centering 
\caption{Performance comparison of different algorithms with QP, 2k-bus system.}
\label{table:QP_2000}
\begin{tabular}{|c|c|c|c|}
\hline
Algorithm & t(s)-CPU & It & Gap(\%) \\
\hline
Adam & 0.68 & 88 & 0.6 \\
\hline
AdaGrad & 0.77 & 107 & 1.2 \\
\hline
GDM & 0.06 & 29 & 1.27 \\
\hline
Gurobi & 0.43 & 16 & - \\
\hline
MOSEK & 1.62 & 28 & - \\
\hline
\end{tabular}
\end{table}



\begin{table}[!bt]
\centering 
\caption{Performance comparison of different algorithms with QP, 4601-bus system.}
\label{table:QP_4601}
\begin{tabular}{|c|c|c|c|}
\hline
Algorithm & t(s)-CPU & It & Gap(\%) \\
\hline
Adam & 0.83 & 156 & 0.2 \\
\hline
AdaGrad & 0.34 & 41 & 3.8 \\
\hline
GDM & 0.66 & 72 & 4.4 \\
\hline
Gurobi & 1.24 & 19 & - \\
\hline
MOSEK & 1.49 & 20 & - \\
\hline
\end{tabular}
\end{table}

\begin{table}[]
\centering 
\caption{Performance comparison of different algorithms with QP for the 10k-bus system.}
\label{table:QP_10000}
\begin{tabular}{|c|c|c|c|}
\hline
Algorithm & t(s)-CPU & It & Gap(\%) \\
\hline
Adam & 5.60 & 88 & 0.44 \\
\hline
AdaGrad & 5.77 & 128 & 1.2 \\
\hline
GDM & 6.78 & 129 & 0.9 \\
\hline
Gurobi & 10.03 & 25 & - \\
\hline
MOSEK & 15.68 & 55 & - \\
\hline
\end{tabular}
\end{table}



All code for this paper is available online in \cite{Crafiei_github_HICSS:2025}.

\section{Conclusion}\label{sec:conclusion}

DCOPF is a crucial problem in many of the power system's core operations and decision-making procedures. The speed of the problem becomes significantly important when, for a single decision on the grid, we need to solve hundreds of separate DCOPF problems simultaneously. In this paper, we present a platform that not only makes each problem solve faster for huge systems but also enables parallelization of the process for more complex operations on the grid. In this paper, GBO has proven to be beneficial for solving DCOPF.
when dealing with small systems, GDM is shown to be the fastest and smoothest method to use, while as the dimensions of the system grow, Adam seems to be a better choice. Our methods in this paper provided tight lower bounds for a single DCOPF problem which were at best 5.43x faster than Gurobi 10. Also, the use of GPU for gradient calculation made each problem solve at best 25\% faster than CPU for the 10k system; GPUs will enable the solving of many DCOPF problems in parallel, which will be the subject of future work.


\section*{Appendices}

\subsection*{A1. Normalization of the input variable}
To canonicalize the DCOPF problems, we normalize the generation variables  to fall in the $[-1,1]$
interval. Given a vector $x$ with upper and lower limits $\overline x$, $\underline x$, we define the normalized generation vector $\tilde x$ via
\begin{subequations}
\begin{align}
\tilde{x} & =\frac{x-\frac{1}{2}(\overline{x}+\underline{x})}{\frac{1}{2}(\overline{x}-\underline{x})}\\
 & =M_{\sigma}^{-1}(x-m_{\mu}),
\end{align}
\end{subequations}
with a mapping back to the non-normalized vector via
\begin{align}
    x=M_{\sigma}\tilde{x}+m_{\mu}.
\end{align}
The normalized input variable $\tilde x$ is now bounded within the unit $\ell_{\infty}$ ball ($B_{\infty}$): 
\begin{equation}\label{sec:normalization}
B_{\infty} = \{{ \tilde x} \in \mathbb{R}^n : \|{ \tilde x}\|_{\infty} \leq 1 \}
\end{equation}
To normalize an LP, we replace all non-normalized primals with their normalized counterparts. That is, we go from this LP:
\begin{subequations}
\begin{align}\underset{x}{\min}\quad & c^{T}x\\
{\rm s.t.}\quad & Ax-b=0\\
 & Cx-d\le0\\
 & \underline{x}\le x\le\overline{x}
\end{align}
\end{subequations}
to this one:
\begin{subequations}
\begin{align}
\underset{x}{\min}\quad & c^{T}\left(M_{\sigma}\widetilde{x}+m_{\mu}\right)\\
{\rm s.t.}\quad & A\left(M_{\sigma}\widetilde{x}+m_{\mu}\right)-b=0\\
 & C\left(M_{\sigma}\widetilde{x}+m_{\mu}\right)-d\le0\\
 & -1\le x\le1,
\end{align}
\end{subequations}
whose primal variables are now normalized.

\subsection*{A2. Dual Norm Definition}
As stated in~\cite{GPU-acc},
the dual norm
of vector $y$, denoted by $\left\Vert y\right\Vert _{*}$, is
given by 
\begin{align}
\left\Vert y\right\Vert _{*}=\max_{\left\Vert x\right\Vert \le1}\;x^{T}y,
\end{align}
where $\left\Vert \cdot\right\Vert$ in the constraint is a properly defined norm. Notably, the $\ell_{p}$ and $\ell_{q}$ norms are duals of each other when
$p$ and $q$ are Holder conjugates (i.e., when $p,q\in[1,\infty]$, and $1/p+1/q=1$).
From ~\cite{GPU-acc}, we have 
\begin{align}
\left\Vert y\right\Vert _{q}=\max_{\left\Vert x\right\Vert _{p}\le1}x^{T}y,\,\forall p,q\in[1,\infty],\;\tfrac{1}{p}+\tfrac{1}{q}=1.\label{eq: pq_conj}
\end{align}

\subsection*{A3. Dual Cone Definition}
The dual cone $K_*$ associated with a cone $K$ is defined as~\cite{MOSEK}
\begin{align}
 K_* = \{ y : y^T x \geq 0 \text{ for all } x \in K \}.
\end{align}
The rotated second-order cone (RSOC) in $\mathbb{R}^{n+2}$ is defined as
\begin{align}
K^{\text{RSOC}} =&\\
\{ (x, y, t) \in \mathbb{R}^n& \!\times\! \mathbb{R} \!\times\! \mathbb{R} : 2xy \geq \|t\|_2^2, x \geq 0, y \geq 0 \}.\nonumber
\end{align}
\begin{align}
K^{\text{RSOC}}  =&\\
\{ (c, b, a) \in \mathbb{R} \!\times\! \mathbb{R} \!\times\! \mathbb{R}^n&  : 2cb \geq \|a\|_2^2, x \geq 0, y \geq 0 \}.\nonumber
\end{align}
These definitions are used in \eqref{eq:socp-1}. Since $K^{\rm RSOC}$ is self-dual, (i.e., $K^{\rm RSOC}$=${K_*}^{\rm RSOC}$) the dual variables associated with the SOC constraints must also fall in the rotated second order cone~\cite{boyd2004convex}. 

 \bibliographystyle{ieeetr}
\bibliography{references}

\end{document}